\documentclass[usenatbib]{mn2e}

\usepackage{graphicx}
\usepackage[space]{grffile}
\usepackage{latexsym}
\usepackage{amsfonts,amsmath,amssymb}
\usepackage{url}
\usepackage[utf8]{inputenc}
\usepackage{hyperref}
\hypersetup{colorlinks=false,pdfborder={0 0 0}}
\usepackage{textcomp}
\usepackage{longtable}
\usepackage{multirow,booktabs}
\usepackage{hhline}

\usepackage{natbib}

\newcommand{\refresponsestart}{}
\newcommand{\refresponseend}{}
\newcommand{\lcdm}{$\Lambda$CDM}%

\title[Pairwise velocities in the ``Running FLRW'' cosmological model]{Pairwise velocities in the ``Running FLRW'' cosmological model}

\author[Antonio Bibiano]{Antonio Bibiano$^1$, Darren J. Croton$^1$\\
$^1$ Centre for Astrophysics and Supercomputing, Swinburne University of Technology, Hawthorn, Victoria 3122, Australia}

\begin{document}

\maketitle

\label{firstpage}

\begin{abstract}
We present an analysis of the pairwise velocity statistics from a suite of cosmological N-body simulations describing the ``Running Friedmann-Lema\^itre-Robertson-Walker'' (R-FLRW) cosmological model. This model is based on quantum field theory in a curved space-time and extends $\Lambda$CDM with a time-evolving vacuum energy density, $\rho_\Lambda$. To enforce local conservation of matter a time-evolving gravitational coupling is also included. 
Our results constitute the first study of velocities in the R-FLRW cosmology, and we also compare with other dark energy simulations suites, repeating the same analysis. We find a strong degeneracy between the pairwise velocity and $\sigma_8$ at $z=0$ for almost all scenarios considered, which remains even when we look back to epochs as early as $z=2$.
\refresponsestart{}We also investigate various Coupled Dark Energy models, some of which show minimal degeneracy, and reveal interesting deviations from \lcdm{} which could be readily exploited by future cosmological observations to test and further constrain our understanding of dark energy. \refresponseend{}
\\
\\
\\

\end{abstract}

\begin{keywords}
Methods: numerical -- Cosmology: theory -- large-scale structure of Universe -- dark energy
\end{keywords}

\bibliographystyle{mn2e_eprint}
\providecommand{\eprint}[1]{\href{http://arxiv.org/abs/#1}{arXiv:#1}}
\providecommand{\adsurl}[1]{}

\providecommand{\ISBN}[1]{\href{http://cosmologist.info/ISBN/#1}{ISBN: #1}}

\section{Introduction}
The field of cosmology has produced a considerable amount of observations which together are remarkably well fitted by a simple 6-parameter model, commonly called $\Lambda$ Cold Dark Matter ($\Lambda$CDM) model. This model's predictions span different types of observables, ranging from the recent Planck mission's Cosmic Microwave Background Radiation (CMBR) measurements \citep{Ade:2015xua}, the distance measurements for Type Ia supernovae \citep{panstarr}, and the distribution of galaxies in the large-scale structure of the Universe (see e.g. \citealt{wigglez}, \citealt{bosspaper}). Fitting these observations in turn has produced very stringent constraints on the model's parameters,  in particular on the abundances of the components of the Universe, including the so called dark energy. 

According to these constraints dark energy accounts for about 70\% of the energy density in today's Universe. From a theoretical point of view it is modelled by adding a constant, $\Lambda$, to the standard equations of general relativity, and is interpreted as an additional component alongside matter and radiation. The presence of dark energy is necessary to account for the accelerated expansion of the Universe  \citep{carrollCC}, and while its treatment is very simple, $\Lambda$ gives us little insight into the nature of the additional component, a situation which astronomers and physicists alike find highly unsatisfactory.

The canonical interpretation of the cosmological constant $\Lambda$ is that of a manifestation of the energy density of the vacuum. This interpretation is very natural but leads to a catastrophic discrepancy between the measured value of $\Lambda$ and the value calculated using any known quantum field theory mechanism \citep{weinbergCC}. This failure to find a simple explanation within the standard cosmological model has motivated a whole new area of research aimed at explaining its nature and behaviour.

Many mechanisms have been developed to understand dark energy while also preserving the successes of the $\Lambda$CDM model. Some treat dark energy as the manifestation of one or more additional fields, as in the various quintessence models \citep{Wetterich88, RatraPeebles}. Others describe its effects in an approximate way by adding degrees of freedom to $\Lambda$CDM, as in the Chevallier--Polarski--Linder \citep{CPLcp, CPLl} and other parametrized models.  Other still rely on modifying Einstein equations of general relativity to obtain the accelerated expansion without the need for additional components, for example the $f(R)$-gravity \citep{HuSawicki} and Galileon family of models \citep{galileon}. 

Regardless of the nature of a particular model it is important that it can make predictions about the observable properties of the Universe. For some observations, like the CMBR and supernovae, a linear treatment is often enough, while for large-scale structure and peculiar velocity measurements, where the linear approximation breaks down, it is necessary to come up with a different strategy.

In the past decade a new branch of cosmology has developed to deal with these regimes using cosmological N-body simulations. These simulations allow us to study the evolution of the matter distribution in the Universe under the influence of both gravity and cosmic expansion down to small scales. To date significant effort has been put into the development of algorithms and highly efficient codes necessary to carry out such simulations with the goal of reaching high precision and large dynamical range. However only recently has the attention shifted towards adapting these algorithms and codes to simulate different cosmological scenarios. Many such modifications have already been carried out for various dark energy models, e.g. the works by \cite{baldiCodecs}, \cite{LiQuinte}, and the modified gravity models as described in  \cite{ecosmog}, \cite{mggadget}, \refresponsestart{}\cite{Oyaizu_2008} and \cite{Llinares_2013}.\refresponseend{}

In this paper we will focus on a scenario that falls partway between the modification categories introduced earlier: the ``Running Friedmann-Lema{\^i}tre-Robertson-Walker'' (R-FLRW) cosmological model. This model was introduced in \cite{Grande_2011} and retains the interpretation of the cosmological constant as vacuum energy but allows for the evolution of its energy density. This in turn requires a variation of the gravitational constant with time to enforce the local conservation of matter. 

Our investigation is motivated by the fact that future surveys, like Euclid \citep{EuclidPap} and eROSITA \citep{eROSITA}, will be able to highlight even small deviations from $\Lambda$CDM, and will potentially be able to distinguish between different but very close cosmological scenarios. 
More importantly, the recent advances of peculiar velocity surveys like SDSS-III \citep{pvssdss3surv}, BOSS \citep{pvsbosssurv} and Pan-STARRS1 \citep{pvspanstarsurv}, constitute a new generation of observational efforts not seen since the 90s (see \cite{pvs2dfsurv} and \cite{pvssdsssurv} for two examples). Such surveys carry with them new stimuli for the theoretical exploitation of phase space. These surveys have shown the potential to constrain, with a high degree of precision, many velocity statistics, such as bulk flows, the velocity divergence power spectrum, the pairwise velocity distribution, and others, which can in turn be compared to predictions from different modified gravity and dark energy models. This provides new ways of testing such models and $\Lambda$CDM.

In this paper we use the suite of N-body cosmological simulations described in \cite{Bibiano} to investigate how the variation of $G$ and $\Lambda$ in the R-FLRW model impacts the clustering of matter.
In the previous work we focused on the differences in the structure formation process by comparing the number counts and the intrinsic properties of haloes and sub haloes across the various scenarios. Here we investigate the changes exhibited in various clustering statistics, considering the full phase-space information obtained from our suite of simulations. More specifically, we will compare the distribution of structures giving particular attention to their relative velocities and how those are affected by the new dynamics.

Our approach will be purely statistical. We will start by introducing the statistics of clustering that underlies our understanding of the spatial nature of the galaxy distribution.\
This theory builds on the galaxy correlation function and the simple clustering pattern it predicts, which in turn lends itself to robust observational measurements.
The dynamics of clustering are theoretically described using the so called Bogoliubov-Born-Green-Kirkwood-Yvon (BBGKY) hierarchy which is borrowed from plasma physics. It allows astronomers to draw relations between the correlation functions of various orders, effectively enabling one to describe how the distribution of every matter particle or particle group can be affected by the perturbations and  evolution of its neighbours.

We will see that this description allows us to predict the behavior of the velocity distribution in the linear regime, which will later be verified in all the R-FLRW scenarios considered.
But the most valuable use of our simulations is their ability to investigate such statistics on mildly non-linear to fully non-linear scales. 
We will highlight the differences on these scales that are inherent to the R-FLRW cosmology. These differences, while measured against both the dark matter distribution and the structure hierarchy within it, bear, under certain assumptions, a direct relation with the galaxy population in the observed Universe.

\ 

\refresponsestart{}This paper is organized as follows: In Section \ref{sec:model} we will introduce the R-FLRW model and the simulation suite. In Section \ref{sec:vel} we will describe the pairwise velocity statistics and their measurement from simulations. In Section \ref{sec:results} we will discuss the results of our measurement. Finally, in Section \ref{sec:discuss} we compare these results to the current literature and summarize.\refresponseend{}

\section{Simulating the R-FLRW scenario}
\subsection{The Model}
\label{sec:model}
The ``Running Friedmann-Lema{\^i}tre-Robertson-Walker'' (R-FLRW) cosmological model was first described in \cite{Grande_2011}. In this model the behaviour of dark energy is explained by quantum field theory calculations on a curved space-time which result in an effective quantity whose value evolves with the expansion of the Universe. No interaction is prescribed between dark energy and the matter component in the Universe as conservation of matter is enforced by allowing for an evolution in the strength of the gravitational coupling. With this additions, the R-FLRW model retains the standard $\Lambda$CDM model's interpretation of $\Lambda$ as the result of vacuum energy, while considering the reasonable possibility that its energy density might be related to other time-varying cosmological quantities. This idea has solid roots in fundamental physics, and we refer the reader to the aforementioned literature for a thorough description of the underlying quantum field theory background necessary \refresponsestart{}to justify some of the model choices\refresponseend{} here. In the present work we will limit our discussion to an introduction of the main equations and the notation necessary for our analysis. More detail can be found in  \cite{Grande_2011}.
\

The R-FLRW model shares the same framework as \lcdm{}: space-time is described using a spatially flat Friedmann-Lema{\"i}tre-Robertson-Walker (FLRW) metric and its interaction with the matter contents of the Universe is described by the Einstein equations.
The cosmological constant term $\Lambda$ is still considered associated with a vacuum energy density $\rho_\Lambda$ which is allowed to evolve with the expansion of the Universe according to the law
\begin{equation}
\label{secondlambdalaw}
    \rho_\Lambda(H) = \rho_\Lambda^0 + \frac{3 \nu}{8 \pi} M_P^2 (H^2 - H_0^2) ~,
\end{equation}
where $H_0$ \refresponsestart{}and $\rho_\Lambda^0$ are the present day values\refresponseend{} of the Hubble parameter and vacuum energy density, $M_P$ is the Planck mass, and $\nu$ is a free parameter which determines the strength of the time variation. Although this form is purely phenomenological, the time scale for the variation is chosen to be $H$. In this way we associate the running of the cosmological quantities to the typical energy scale of the gravitational field associated with the FLRW metric.

It is worth emphasising that the parameter $\nu$ is a critical component of the new framework; when $\nu = 0$ the vacuum energy remains constant with $\rho_\Lambda = \rho_\Lambda^0$, and the model reduces to \lcdm{}. In \cite{Grande_2011} $\nu$ was considered a free parameter with a natural range of $|\nu| \ll 1$. More specifically, the range of $\nu$ was constrained against joint supernovae, CMBR and BAO observations to lie in the range $-0.004 < \nu < 0.002$. This ensures the R-FLRW model is consistent with current cosmological measurements at least at the 1$\sigma$ level.

The possibility of a running of the cosmological constant was studied in \cite{lambdaonly}, but in the scenarios considered here the local conservation of matter is additionally enforced by allowing the gravitational coupling constant $G$ to also evolve with the expansion of the Universe. The evolution law for $G$ can be obtained through the Bianchi Identites \citep{Grande_2011}, which result in the following logarithmic form:
\begin{equation}
\label{variationofg}
    g(H) \equiv \frac{G(H)}{G_0} = \frac{1}{1+\nu \text{ ln } (H^2/H_0^2)} ~.
\end{equation}

The system of equations describing the background expansion of the Universe in the R-FLRW model, can be written as:
 \begin{align}
    & E^2(z) \equiv H^2(z)/H_0^2 = g(z)[\Omega_m(z) + \Omega_\Lambda(z)] ~, \label{syshubble}\\
    & (\Omega_m + \Omega_\Lambda) ~\text{d}g + g ~\text{d} \Omega_\Lambda = 0 ~, \label{sysbianchi} \\
    & \Omega_\Lambda(z) = \Omega_\Lambda^0 + \nu [E^2(z) - 1] ~, \label{syslambda} \\
    & \Omega_m(z) = \Omega_m^0 (1+z)^{3(1+w_m)} \label{sysom}~.
\end{align}
The first equation is the R-FLRW version of the Friedmann equation in the $\Lambda$CDM model, the second is the differential form of the Bianchi equation, the third is just Equation~\ref{secondlambdalaw} rewritten using the density parameter, and the last is a rewrite of the standard equation for $\rho_m$ generalised to include relativistic ($w_m = \frac{1}{3}$) and non-relativistic ($w_m = 0$) matter. We write this system of equations in terms of the density parameters, $\Omega_m$ and $\Omega_\Lambda$, defined as the energy densities normalized to the critical density, $\rho_c^0 = \frac{3H_0^2}{8 \pi G_0}$, at the current epoch.

\subsection{Perturbations}
\label{sec:perturb}
The linear perturbations for the R-FLRW model were studied in \cite{Grande_2010}, who show how a full mathematical treatment must include the perturbations for $\rho_\Lambda$ and $G$. These in turn influence the matter perturbations according to the following second order differential equation that describes the evolution of the matter density contrast, $\delta_m \equiv \delta\rho_m/\rho_m$:
\begin{equation}
\label{scalefactorfirst}
    \delta_m'' + \left( \frac{3}{a} + \frac{H'}{H} \right) \delta_m' = \frac{3}{2 a^2} \left( \tilde{\Omega}_m \delta_m +  \tilde{\Omega}_\Lambda \delta_\Lambda + \frac{\delta G}{G} \right ) ~.
\end{equation}
Here a prime denotes differentiation with respect to the scale factor, $a$, and the tilde over each density parameter means that the energy density is normalized to the critical density at the same redshift, $\rho_c(z) = \frac{3H(z)^2}{8 \pi G(z)}$.
After some manipulation this equation lends itself to a numerical solution, and in \cite{Bibiano} we showed how it predicts an enhancement of growth when $\nu < 0$, due to the strengthening of the gravitational coupling at high redshift that allows the perturbations to overcome the ``repulsion'' associated with expansion, caused by the vacuum energy density. The converse was evident when $\nu > 0$, where the higher value of $\rho_\Lambda$ and the weakening of the gravitational coupling at high redshift hindered the early growth of perturbations.

This behaviour is the main distinguishing feature of the linear analysis of the model. In fact, and as was demonstrated by \cite{Grande_2010}, the shape of the matter power spectrum is the same for the R-FLRW model as that of $\Lambda$CDM.  This is due to the perturbations in $G$ and $\rho_\Lambda$ being negligible at early times.  This allows us to set the power spectrum amplitude for all models at the CMBR redshift, $z\sim 1100$, to make them fully compatible with the Planck mission's observations \citep{Ade:2015xua}. 

\subsection{Simulations}
\label{sec:simul}

\begin{table}
\begin{center}
\begin{tabular}{lc}
\hhline{==}
Parameter & Value\\
\hline
$\Omega_m$ & 0.3175 \\
$\Omega_\Lambda$ & 0.6825 \\
$\Omega_b$ & 0.0490 \\
$h$ & 0.6711 \\
$n$ & 0.9624 \\
$\sigma_8$ & 0.8344\\
\hhline{==}
\end{tabular}
\caption{Cosmological parameters at redshift $z = 0$ from the Planck mission as reported in \protect\cite{Ade:2015xua}. These have the same definition in both the $\Lambda$CDM and in the R-FLRW models and are used in all our simulations.}
\label{tab:cosmpar}

\end{center}
\end{table}

To explore structure formation in the R-FLRW model we performed a suite of dark matter only N-body simulations.  These simulations follow the evolution of $1024^3$ cold dark matter particles, each of mass $\sim 8\times10^{10} ~M_{\odot}/h$, in a periodic cosmological box of $1024 ~{\rm Mpc}/h$ on a side with 62 snapshots spanning the expansion history from $z=49$ to $z=0$. The suite consists of four simulations covering the natural interval for the $\nu$ parameter as described above from -0.004 to 0.002, and one control simulation that uses the standard $\Lambda$CDM cosmology. The present day cosmological parameters are the same between the simulations and reflect the latest Planck mission \citep{Ade:2015xua} determination for $\Omega_m$, $\Omega_\Lambda$, $\Omega_b$, $h$, $n$ and $\sigma_8$. These values are reported in Table \ref{tab:cosmpar}.

\begin{table}
\begin{center}
\begin{tabular}{lcc}
\hhline{===}
Simulation & $\sigma_8(z = 0)$ & $\nu$ \\
\hline
H1 & 0.8916 & -0.004 \\
H2 & 0.8479 & -0.001 \\
H3 & 0.8215 & +0.001\\
H4 & 0.8090 & +0.002 \\
\hhline{===}
\end{tabular}
\caption{The different R-FLRW simulations performed. The three columns are the simulation name, the value of $\sigma_8$ at redshift $z=0$ reached by each scenario and the value of the $\nu$ parameter of the R-FLRW model simulated.}
\label{tab:sims}
\end{center}
\end{table}

The simulations were carried out using a modified version of the parallel TreePM N-body code \textsc{gadget-3} \citep{Springel_2005}. This version keeps the original algorithms that evolve the dark matter particles but interpolates the cosmological quantities $H(z)$, $G(z)$, $\Omega_m(z)$, and $\Omega_\Lambda(z)$ using look-up tables. This addition is needed because such quantities now evolve differently in a R-FLRW universe compared with the standard $\Lambda$CDM that \textsc{gadget-3} usually assumes. The shift to look-up tables also helps make numerical implementation of the model more manageable and avoids an otherwise inevitable performance hit at every time-step. It also makes the code more general and applicable to other scenarios, such as with our reference \lcdm{} simulation. This avoids any potential differences due to the use of different codebases in the comparison.

Next, in order for the N-body code to correctly calculate the potential for each of the new models, additional changes need to be made. In particular, the effect of the perturbations in $\rho_\Lambda$ and $G$ must be taken into account. To do so in our previous study we introduced the approximation
\begin{equation}\label{eq:ansatz}
\frac{\delta G}{G} = f(a)\frac{\tilde{\Omega}_m}{1-\tilde{\Omega}_\Lambda}\delta_m ~,
\end{equation}
which leads to an additional modification in the formula for the gravitational potential,
\begin{equation}
\phi = - \frac{3}{2} \frac{H^2a^2} {k^2} (1+ f(a)) \tilde{\Omega}_m \delta_m ~.
\end{equation}
This is similar to what is solved by a standard $\Lambda$CDM N-body algorithm, with an additional time dependency through $f(a)$ which can be incorporated in the time dependence of $G(a)$. The functional form of $f(a)$ was constrained using the results from linear perturbation theory and the correction was included in the calculation of the look-up tables. In \cite{Bibiano} we showed that this correction is able to recover the theoretical linear growth of structure to within 0.5\%.

\subsection{Initial Conditions}
\label{sec:initcond}
Finally, the last part of the simulation pipeline that required careful consideration given the change in cosmology was the generation of the initial conditions. These were obtained by perturbing a glass particle distribution according to the 2LPT prescription described by \cite{Crocce_2006} using the 2LPTic code. \refresponsestart{}This code assumes \lcdm{} to calculate the growth rate and the value of the Hubble function at two redshifts: the starting redshift chosen to be $z=49$, and the present time. We instead calculated these values using the look-up tables described above, and substituted the hard coded formulas with two run-time input values.\refresponseend{}

To generate the initial conditions we draw the phases from a random distribution assuming a predefined shape and amplitude of the linear power spectrum. We initialize the distribution using the same random seed across the different scenarios and use the same power spectrum shape obtained with the \textsc{camb} code described in \cite{Lewis:2002ah}. This code calculates the $\Lambda$CDM linear power spectrum shape at any redshift given the cosmological parameters at redshift $z=0$. We use the same calculation for our R-FLRW simulations since, as discussed above, this model retains the same linear power spectrum shape as that of the $\Lambda$CDM model.   The amplitude of the power spectrum is then set according to the $\sigma_8$ value given in Table \ref{tab:cosmpar}, scaled back to the epoch of recombination ($z = 1100$) using the standard $\Lambda$CDM formula for the growth factor, and then scaled forward to the initial redshift of our simulations, $z=49$, using the numerical solution of Equation~\ref{scalefactorfirst}.   This choice is equivalent to normalising the power spectrum of every realisation to the same $\sigma_8$ at the epoch of recombination and results in a different $\sigma_8$ at redshift $z=0$, as reported in Table \ref{tab:sims}. This, as we discussed in Section \ref{sec:perturb}, makes our models compatible with the Planck mission's observations and allows us to focus our analysis on the resulting differences measurable at low redshift.

\subsection{Halo Finding}
Our version of the \textsc{gadget-3} code performs two levels of halo identification for every snapshot saved. These routines inherit the modifications previously discussed, but aside from this were otherwise unchanged. First, the Friends-Of-Friends (FOF) algorithm \citep{Davis_1985} used by \textsc{gadget-3} identifies halos based on a nearest neighbour search, with a linking length $b=0.2$ of the mean inter-particle separation. The mean density of such halos approximately correspond to the overdensity of virialised structures expected from the spherical collapse model. Second, substructures are then traced using the \textsc{subfind} \citep{Springel_2005} algorithm that groups gravitationally self-bound particles around local density maxima so that every FOF group contains at least one sub-halo.

\section{Velocities}
\label{sec:vel}

\subsection{The Pairwise Velocity Distribution}
\label{sec:velintro}
The mean relative velocity of a pair of galaxies was first discussed from a theoretical point of view by \cite{davispeebles}, as a consequence of the Bogoliubov-Born-Green-Kirkwood-Yvon (BBGKY) hierarchy of equations used to describe the dynamics of clustering.\
With this description we can obtain a series of evolution laws for any n-point clustering statistic. In particular, we apply this method to the correlation function, where at every step of the hierarchy, to describe the dynamics of the n-point correlation function we need to know the n+1-point correlation function. Hence, to proceed with the description of a n-particle system usually a physically motivated approximation or a subset of the phase-space must be considered.

This statistical description is particularly useful because in this picture matter is approximated as a collection of identical particles of mass $m$. We can think of these particles as a random sample of the underlying distribution of dark matter and baryons, or, under assumptions that will be explained later, as individual galaxies. This makes the theoretical results derived in this framework akin to the results from N-body simulations whose description of the Universe stems from similar assumptions.

A first consequence of the BBGKY equations is the conservation of particle pairs. In fact, the two point correlation function $\xi$, which describes the clustering of particle pairs, is related to the relative velocity of the two particles in the pair $v$:
\begin{equation}
\label{eq:pairscons1}
\frac{\partial \xi}{\partial t} + \frac{\partial}{\partial x^i }\left [\frac{v}{a} \frac{x^i}{x}(1+\xi)\right ] = \frac{\partial \xi}{\partial t} + \frac{1}{x^2a}\frac{\partial}{\partial x }[x^2 (1+\xi)v] = 0,
\end{equation}
where $v$ is the projection of the 3D velocity vector onto the separation $\vec{x}$:
\begin{equation}
\vec{v} \cdot \vec{x}  \longleftrightarrow v^i = v \frac{x^i}{x}.
\end{equation}
If we consider just the neighbors of a particle within a distance $r$ we can express Equation \ref{eq:pairscons1} in integral form:
\begin{equation}
\label{eq:pariscons2}
\frac{\partial}{\partial t} \left( na^3 \int_0^r dx ~ 4\pi x^2 [1+\xi(x,t)] \right) = -4\pi a^2 x^2 n(1+\xi) v .
\end{equation}
Here $n$ is the number density of particles and this expression makes very clear the conservation of particle pairs when we interpret the right hand side as the mean flux of neighbors out of a surface with $x$ = constant, and the left hand side as the rate of change of the mean number of neighbors within a distance $r$ from the particle.

When solving Equation \ref{eq:pariscons2} we can identify two regimes characterized by different scales.
On small scales, where the dynamics are dominated by the gravitational interaction within clusters, we expect the relative velocities to average around zero. Therefore, $v$ should only change due to the Universe's expansion according to  the Hubble law:
\begin{equation}
v = -\dot{a}x ~.
\end{equation}
In the literature this is known as the ``stable clustering regime''.

On the other hand, on large scales, where the clustering still follows a linear prescription, we can use the fact that the correlation function averaged over a volume with comoving radius $x$ is proportional to the growing mode of the linear density contrast $D_+$:
\begin{equation}
\bar{\xi}(x,t) \equiv \int_0^x y^2 dy \xi(y,t) \propto D_+^2(t)~.
\end{equation}
If we plug this last relation into the pair conservation Equation \ref{eq:pariscons2} we obtain
\begin{equation}
v = - \frac{2a}{x^2 (1+\xi)} g(t) \int_0^x dx x^2 \xi(x,t) ~,
\end{equation}
where we have defined the growth rate $g(t) \equiv \dot{D}_+/D_+$.

Is it also possible to interpolate between the linear and stable clustering regime solutions, as shown by \cite{pwspap1} using the parametrized approximation
\begin{equation}
v = -\frac{2}{3} H x g \bar{\bar{\xi}} (x) (1+\alpha  \bar{\bar{\xi}}(x)) ~,
\end{equation}
where $\bar{\bar{\xi}} =  \bar{\xi}/x^3(1+\xi)$ and $\alpha$ is a parameter related to the slope of the correlation function.

One feature that all these analytical expressions have in common is that the pairwise velocity has a strong dependence on both the growth rate of matter perturbations and on the  clustering of matter, and this makes it an excellent candidate to distinguish between the various R-FLRW scenarios and $\Lambda$CDM.

\subsection{Measuring From Simulations}
\label{sec:meassim}
The post processing products of our simulations suite can be exploited to explore how the different R-FLRW scenarios impact the pairwise velocity distribution.
To do so we first need to measure the magnitude of the radial pairwise velocity for all pairs of objects. An object here can be either a particle or a halo, and in the next section we will perform measurements for both categories. The first directly probes the matter distribution, while the second has a closer connection to observations.

The pairwise velocity, defined as the velocity difference $\vec{v}_1 - \vec{v}_2$ for a pair of objects, can be separated into two components. The first is the projection onto the separation vector $\vec{r}$, 
\begin{equation}
\label{eq:vonetwofromsims}
\vec{v}_{12} = (\vec{v}_1 - \vec{v}_2) \cdot \frac{\vec{r}}{|\vec{r}|} ~,
\end{equation}
and the second is perpendicular to $\vec{r}$,
\begin{equation}
\vec{v}_\perp = (\vec{v}_1 - \vec{v}_2) - \vec{v}_{12} ~.
\end{equation}
From this we can measure their averages and dispersions, $\sigma_{||}^2 = \langle v_{12}^2 \rangle$ and $\sigma_{\perp}^2 = \langle v_{\perp}^2 \rangle$ respectively.

While these quantities are readily available in our simulations they are not directly observable. But it is possible to connect them to observations by means of estimators like the ones described in \cite{ivarsen}. Using these it is possible to estimate $v_{12}$ and $\sigma_{||}$ with only line-of-sight velocity and separations. Another approach is to measure line-of-sight and projected quantities directly  from the simulations and bin them in intervals of transverse and line-of-sight separation, to calculate the so-called projected line-of-sight velocity dispersion,
\begin{equation}
\sigma^2_p(r_p,\pi) = \frac{r_p^2v_\perp^2/2+\pi^2(v_{||}^2-v_{21}^2)}{r_p^2+\pi^2} ~,
\end{equation}
where $r_p$ is the projected separation and $\pi$ is the line-of-sight separation.
We can then integrate this quantity along the line of sight to obtain the line-of-sight velocity dispersion,
\begin{equation}
\label{eq:sigma12def}
\sigma^2_{12}(r_p) \equiv \frac{\int dl \xi(r) \sigma^2_p(r)}{\int dl \xi(r)}.
\end{equation}
According to \cite{Jenkins:1997en}, this quantity is closer to those measured in galaxy redshift surveys and can be readily compared to observations. 

To obtain both the line-of-sight and the projected separation from our simulations we will use the distant observer approximation. Here an observer is assumed to be positioned at infinity along the $z$ direction that becomes its line-of-sight. In this way the $z$ component of the physical separation $r$ becomes $\pi$ and the component in the $x$-$y$ plane becomes $r_p$.

To perform the measurement on a simulation snapshot one needs to consider all possible pairs of objects and calculate for each pair the distance and all the velocity statistics explained above. This quickly becomes computationally unfeasible as it scales with the square of the number of objects. Since the pairwise velocity drops to zero on large scales, to speed up the calculation we can leverage the publicly available Corrfunc code of \cite{manodeepcode}. This code allows one to reduce the computational cost of calculating the correlation function for a given set of objects by fixing the maximum distance to which the correlation function is computed. We extended this code to compute the pairwise statistics needed for this work.\footnote{The modification have been made publicly available at \url{https://github.com/antbbn/Corrfunc} .}

\section{Results}
\label{sec:results}

\begin{figure}
\includegraphics[width=\columnwidth]{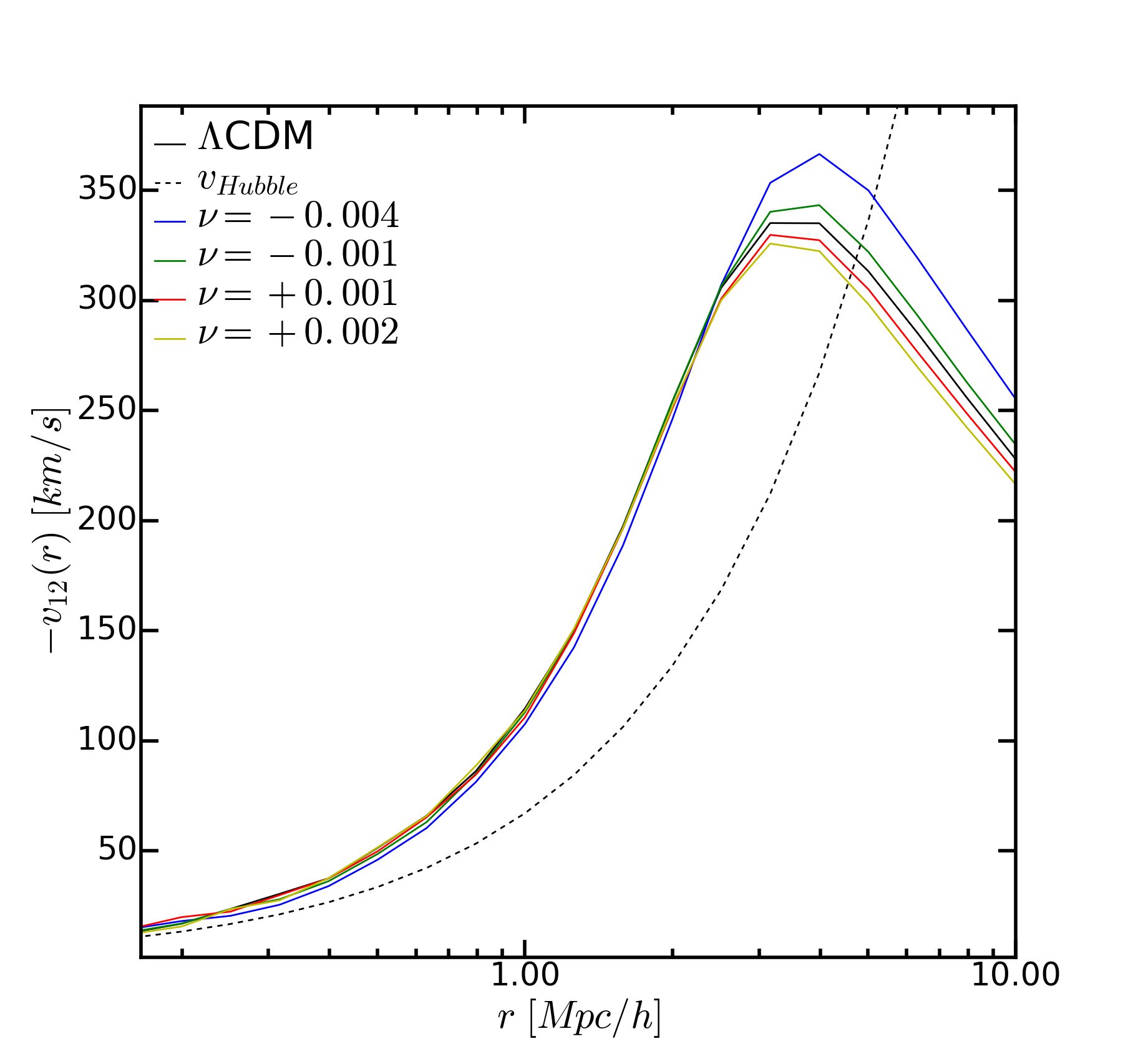}

\caption{ The mean inward radial pairwise velocity, $v_{12}$, as a function of the physical separation $r$. This is measured using a representative sample of dark matter particles from our simulation suite. The dashed line is the Hubble velocity given by $v_{Hubble} = -Hr$, where $H$ is the Hubble constant. }
\label{fig:vonetwoparticles}
\end{figure}

\subsection{Dark Matter Distribution}

The first result that we show is the mean pairwise velocity for particle pairs as a function of physical separation $r$.
\refresponsestart{}To make this calculation manageable we measure $v_{12}$ for a random subset of the simulation particles which contains around 20\% of the total simulation particles, and then collect the results in equally spaced bins in $\log(r)$.\refresponseend{} We repeat this procedure for the five simulations in our suite and plot the outcome in Figure \ref{fig:vonetwoparticles} alongside the Hubble velocity.

For all the simulations $v_{12}$ follows the Hubble line up to $\sim$ 300 kpc/h. This means that the R-FLRW modifications do not influence the scales at which structures are fully relaxed. For larger separations, where $v_{12}$ exceeds the Hubble velocity, the simulations show again little deviation from each other up to 4 Mpc/h, where the mean infall velocity reaches a peak. 
Here the models do not exhibit any clear difference in the position of the velocity peak but do show a change in its magnitude, with the $\nu < 0$ R-FLRW scenarios resulting in larger infall velocities and the $\nu > 0$ scenarios resulting in smaller velocities, when compared to $\Lambda$CDM. The $\nu = -0.004$ simulation shows the most significant deviation, with a peak velocity about 50 km/s higher than $\Lambda$CDM.
The difference in the peak leads also to differences in the separation at which the crossover between the pairwise velocity and the Hubble velocity occurs, which marks the typical scale at which motions inside structure become closer to isotropic.
This scale is larger for $\nu < 0$ and smaller for $\nu > 0$, and only different by a few kpc between all the R-FLRW scenarios considered.

\subsection{Dark Matter Halos}
\label{sec:dmhalosvel}
Measuring the pairwise velocity distribution for pairs of particles is tantamount to a measurement on the overall dark matter distribution. And while this does not correspond to properties that we are able to observe, we expect these features to be mirrored in the distribution of virialized objects. For this reason we now switch our focus to dark matter halos and examine the pairwise velocity statistics introduced in Section \ref{sec:meassim} using the halo catalogs obtained in the post-processing phase of our simulations. \refresponsestart{}We perform these measurements separately for different halo mass ranges, considering only halo pairs within the same mass range, this will allow us to also highlight differences between the halo populations.\refresponseend{}
\begin{figure*}
\includegraphics[width=\columnwidth]{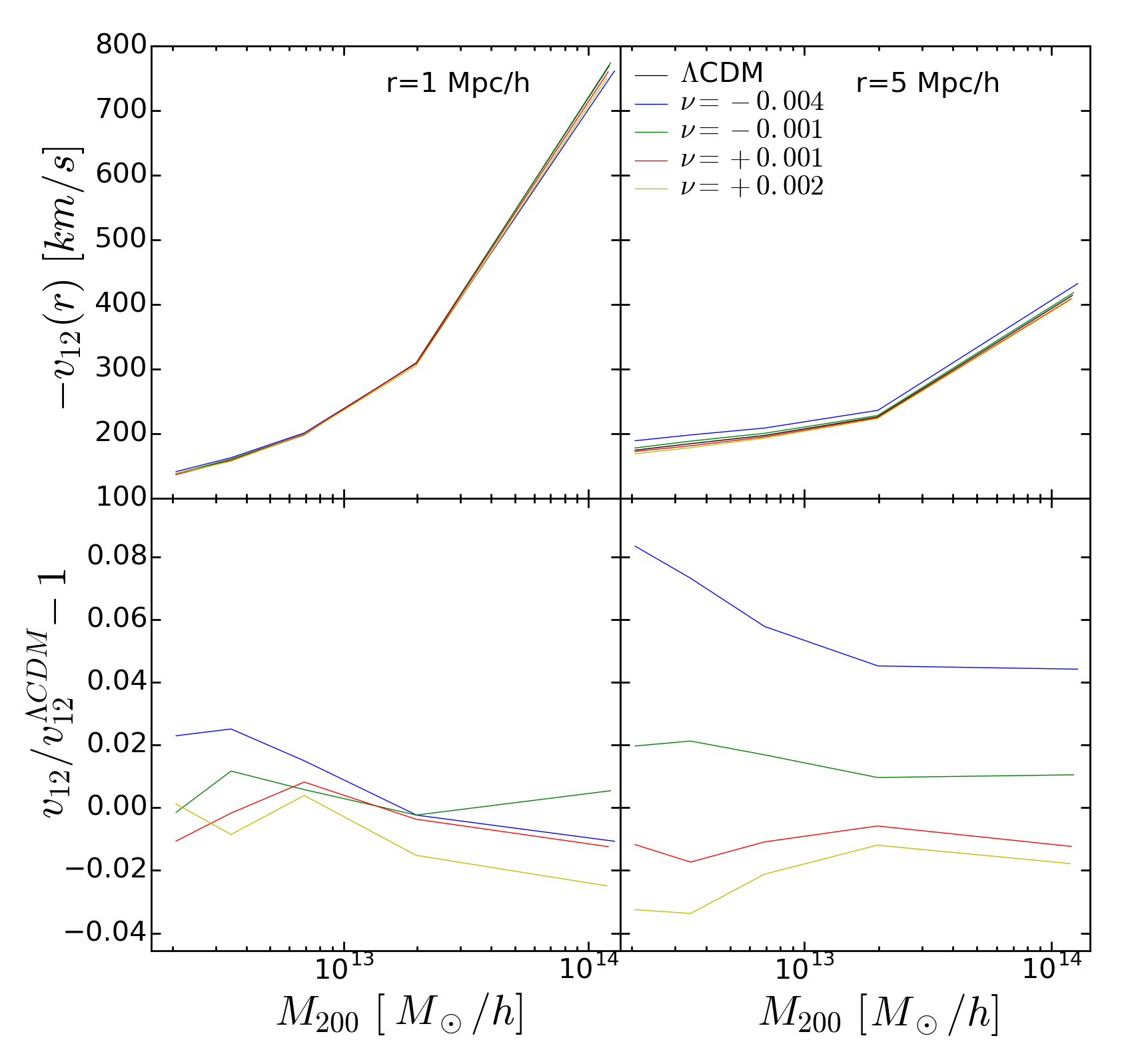}
\includegraphics[width=\columnwidth]{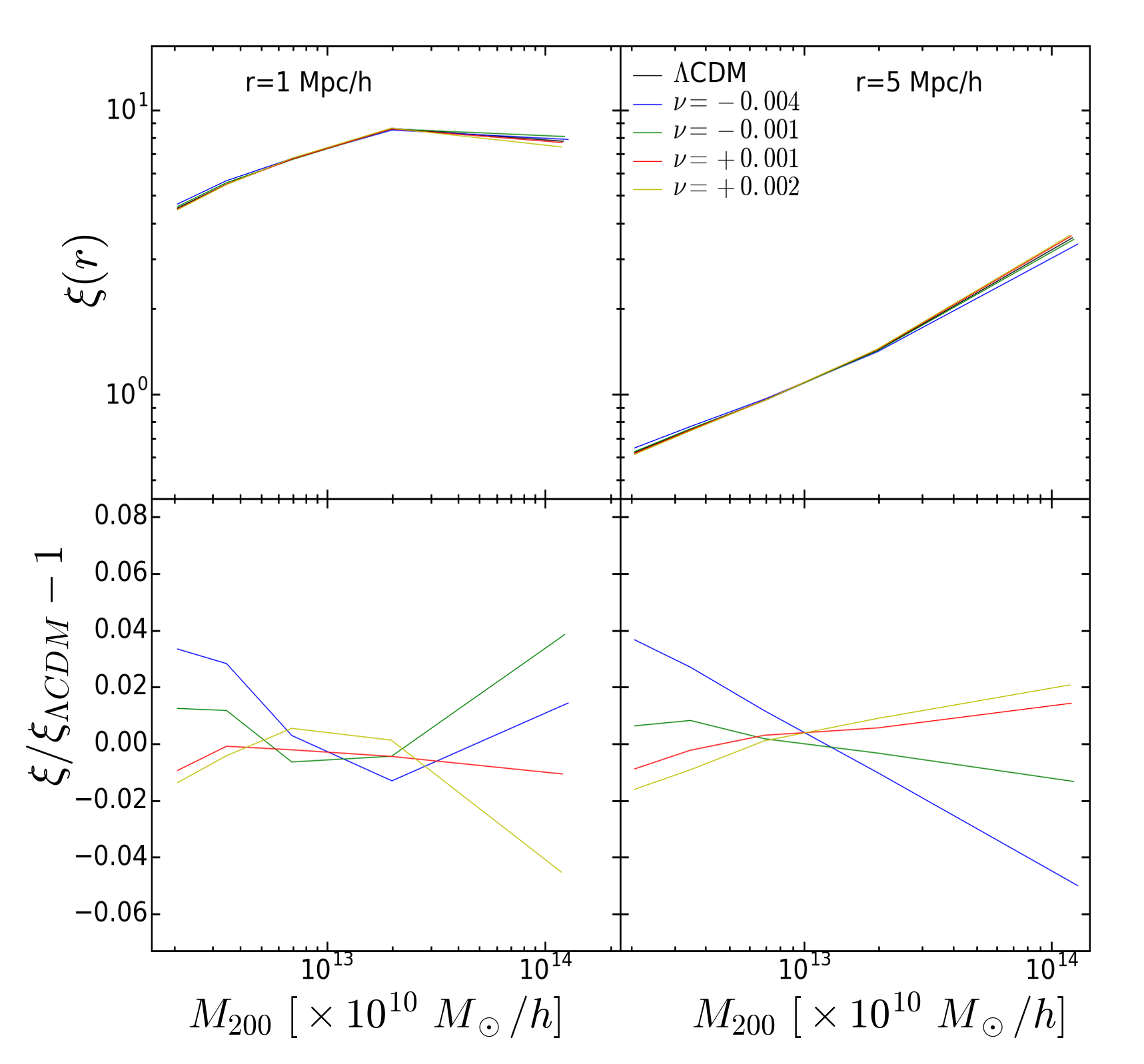}
\includegraphics[width=\columnwidth]{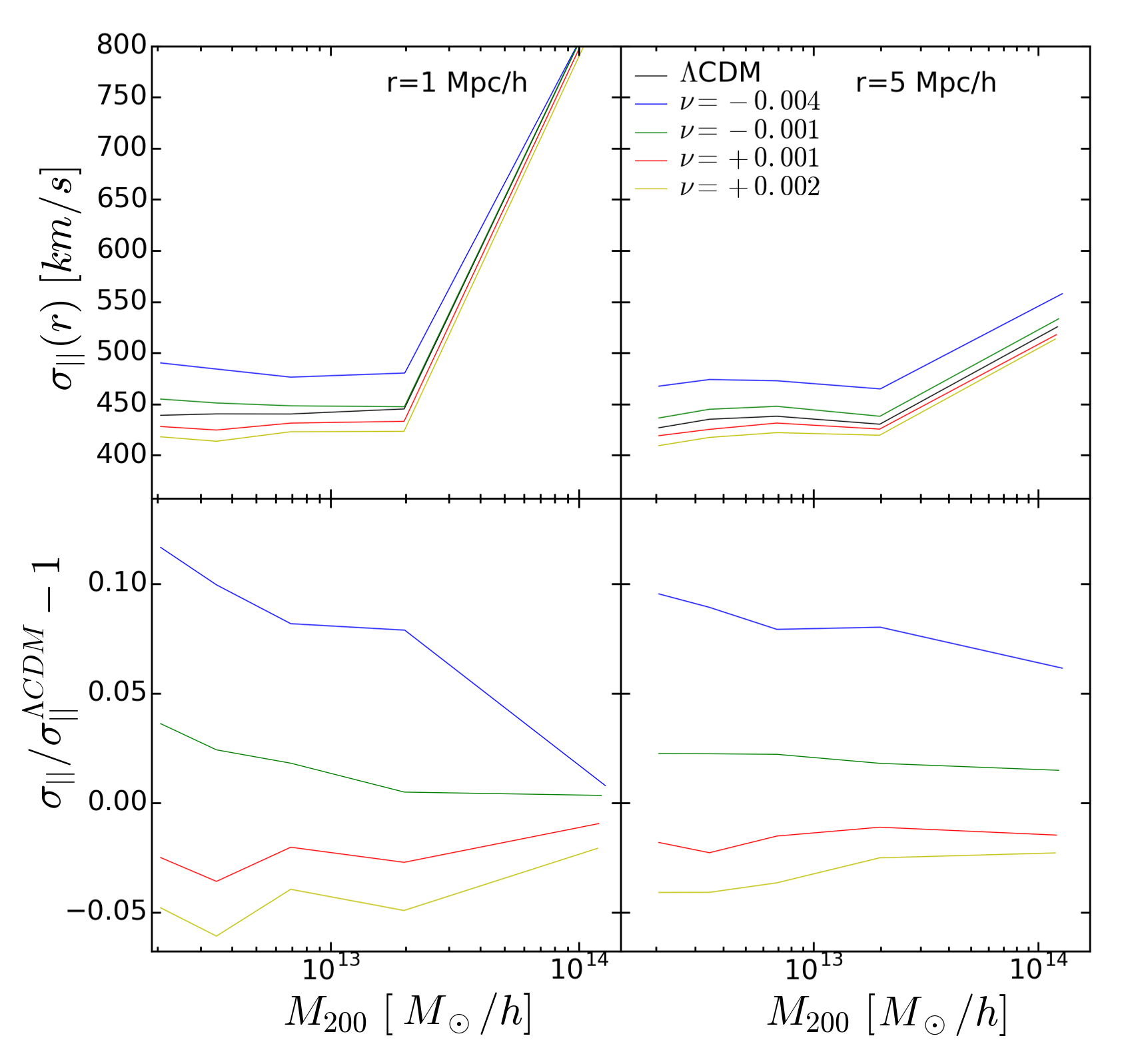}
\includegraphics[width=\columnwidth]{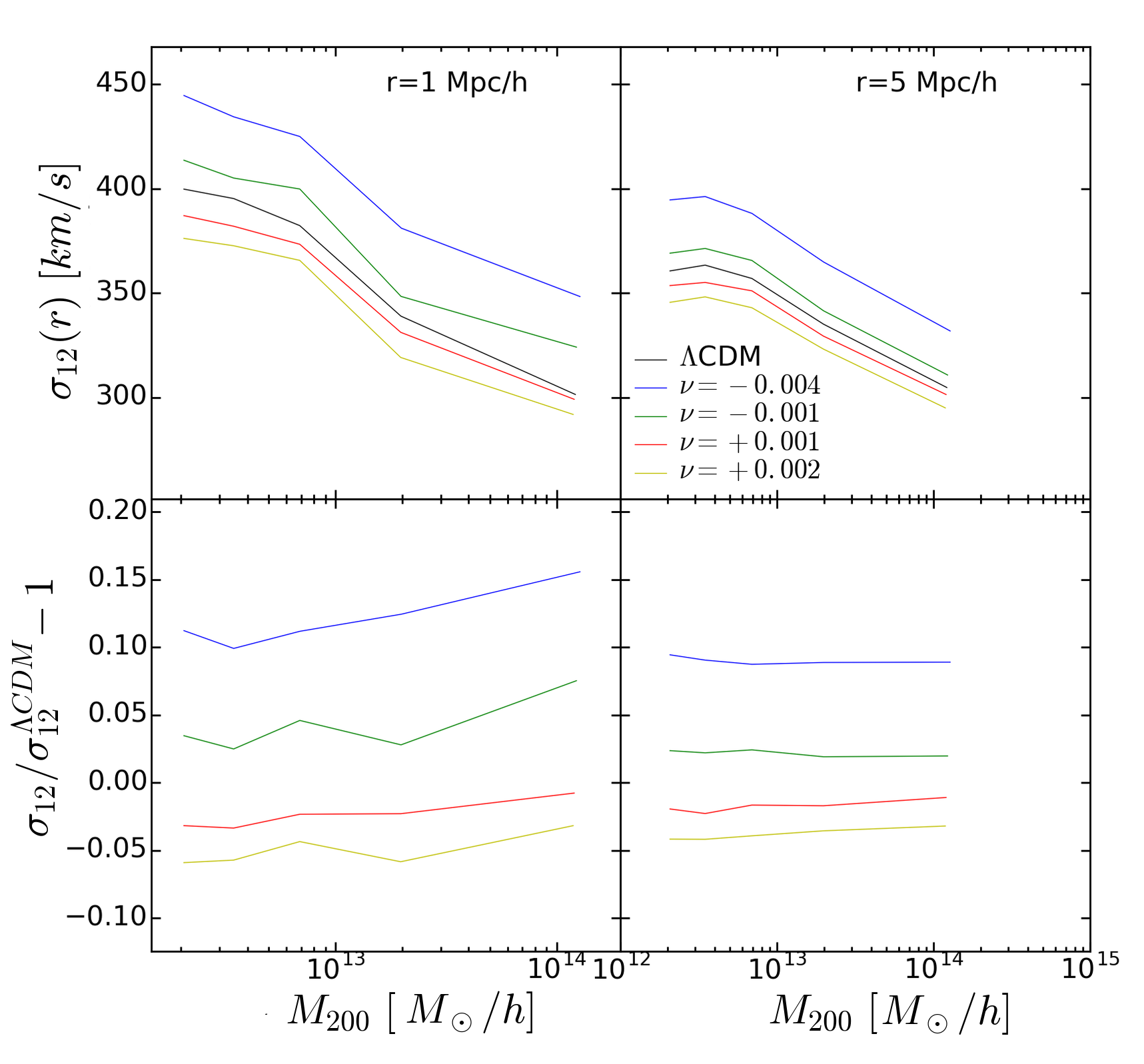}

\caption{ Clockwise from the top left: The mean inward radial pairwise velocity, $v_{12}$, the correlation function, $\xi(r)$, the mean dispersion in the radial pairwise velocity, $\sigma_{||}$ and the line of sight dispersion, $\sigma_{12}$, all as a function of halo mass for two physical separations, $r = 1$ Mpc/h and $r = 5$ Mpc/h, for all the simulations in our suite. \refresponsestart{}The mass is the average mass in the five mass bins in which we split the halo catalog. Pairs are restricted to halos within the same halo mass bin, as described in Section~\ref{sec:dmhalosvel}.\refresponseend{}~
 In the bottom panels we plot the residuals with respect to the reference $\Lambda$CDM simulation.  }
\label{fig:fourplots}
\end{figure*}

The mean radial pairwise velocity for dark matter halo pairs is shown in the top left plot of Figure \ref{fig:fourplots},  for separations of 1 Mpc/h and 5 Mpc/h as marked (Equation \ref{eq:vonetwofromsims}). Across both separations we can see a clear mass dependence, with the velocity increasing with halo mass. This dependence is much steeper at small separations, an effect that is mostly due to the low number of high mass halos that lie this close. The differences between the simulations are much smaller than what was measured for particle pairs. Here, the deviations for all halo masses are well within 10\% of the $\Lambda$CDM case. This is highlighted in the lower panel where the ratio of each R-FLRW scenario to $\Lambda$CDM is plotted. We note that for small separation it is difficult to see a difference between the models. This is mostly due to the mass resolution of our simulations, which fails to recover a large enough number of halos at separations smaller than 1 Mpc/h, but is also expected from the theoretical behavior of $v_{12}$. In fact, as we discussed in Section \ref{sec:velintro}, $v_{12}$ will to converge towards zero at very small separations in such systems. At larger separations we find a shallow mass dependence in the simulation differences, with lower mass halos exhibiting larger deviations than halos of higher mass. 
As expected from theory, low mass halos produce a clear turnover around separations of a few Mpc/h. Again the largest deviations are seen in the $\nu =-0.004$ simulation for the smallest mass halos and largest separations, but rarely exceed 10\%. We omit plots of $v_{12}$ as a function of separation here as they show very similar behavior to Figure \ref{fig:vonetwoparticles} for dark matter particles.

\

A necessary step when calculating pairwise statistics is the measurement of the correlation function, and this allows us to compare clustering as a function of scale and mass across the different R-FLRW scenarios. 
In the top right plot of Figure \ref{fig:fourplots} we plot $\xi(r)$ as a function of halo mass for all our simulations, again taken at two halo separations of $r = 1$ Mpc/h and $r = 5$ Mpc/h. Is it immediately clear that the R-FLRW modifications have only a small impact on halo clustering. In fact, the trends are very similar to the well known trends of \lcdm{}, with the correlation function amplitude increasing with mass and decreasing with separation. While deviations appear at both the high and low mass end, they lie consistently within 5\% of the $\Lambda$CDM realization.  We find the largest differences when $\nu = -0.004$,  with the trends reversing as we consider $\nu > 0$. We note though that the magnitude of these differences might be too small to be observed, even with future surveys.

\

We now move our focus to the dispersions in the pairwise velocity statistic and analyze the radial pairwise velocity dispersion, $\sigma_{||}$ (defined in Equation \ref{eq:vonetwofromsims}).
In the lower left plot of Figure \ref{fig:fourplots} we show $\sigma_{||}$ for all the simulated R-FLRW scenarios as a function of halo mass, using the two separations considered previously. We can see how the dispersion is almost constant across all masses except for the highest mass halos, which have a significantly higher dispersion. We omit a plot of  $\sigma_{||}$ as a function of separation for clarity, but can report it is almost constant across all separations, with a characteristic dip at intermediate separations, which was also found by \cite{hellwing}. \refresponsestart{}At the high mass bins end we find that the dispersion increases rapidly at small separations.\refresponseend{} This is a natural effect of the size of these halos, similar to what we found in the $v_{12}$ measurement.
The differences between the four R-FLRW scenarios and $\Lambda$CDM are larger than the ones measured in the pairwise velocity and reach a maximum for the $\nu = -0.004$ case of a 10\% higher dispersion than $\Lambda$CDM. On the other hand, for  $\nu > 0$ we find a lower dispersion with differences well below 5\%. These differences seem to have almost no mass dependence and a very shallow dependence on the separation, especially for low mass halos.

\

The radial dispersion $\sigma_{||}$, remains inaccessible observationally. For this reason we introduced $\sigma_{12}$ in Equation \ref{eq:sigma12def}.
This quantity is not only closely related to what observers can actually measure, but, by virtue of its definition, also includes various contributions from all the quantities already described. Therefore we expect this observable to display even larger deviations across our mass and separation ranges.
The results from our simulation suite are reported in the lower right plot of Figure \ref{fig:fourplots} and reveal that this is indeed the case: $\sigma_{12}$ has a marked dependence on halo mass, with small halos showing higher dispersion and a similar slope across the two separations pictured. Furthermore, while we again omit the plot of $\sigma_{12}$ as a function of separation for clarity, we find that $\sigma_{12}$ shows a shallow decrease with separation, similar to that displayed by $\sigma_{||}$ and the different mass ranges exhibit a different slope for this dependence, with the dispersion becoming almost constant in the high mass bins. 
As we expected, there are important differences between the R-FLRW and $\Lambda$CDM cosmologies. These differences are nearly constant across all separations and mass ranges, oscillating at around 10\%, with a peak of 15\% for $10^{14} ~ M_\odot/h$ halos in the $\nu =-0.004$ case, and for other scenarios well within 5\%. There is also a clear dependence on the sign of the $\nu$ parameter, with $\nu < 0$ showing higher $\sigma_{12}$ values and $\nu > 0$ showing consistently smaller values when compared to $\Lambda$CDM. 
\

\subsection{Degeneracies}
\label{sec:degeneracies}
As we stressed in Section \ref{sec:initcond}, the choice of initial conditions for our simulations, while allowing us to highlight the low redshift differences between the various scenarios simulated, also makes many of our results degenerate with the value of $\sigma_8$.
For this reason, to complete our analysis we now investigate the degeneracy of our R-FLRW velocity statistics with \lcdm{} scenarios having the same $\sigma_8$.

\

To perform this last step it is necessary to determine the pairwise velocity statistics for various $\Lambda$CDM scenarios which reach the same $\sigma_8$ value at redshift $z = 0$ as the R-FLRW scenarios in our suite. In \cite{Bibiano} this determination was obtained with the aid of fitting functions or computable models from the literature. Here such aids are not available and the only solution is to run additional simulations with the same parameters as our original $\Lambda$CDM simulation but with an adjusted $\sigma_8$. 
To this end we ran two simulations with bounding values of $\sigma_8$ equal to our $\nu = -0.004$ and $\nu = + 0.002$ R-FLRW scenarios; these show the largest deviations across all the observables analyzed and we expect the other scenarios to always fall in between.

\begin{figure*}
\includegraphics[width=1.05\textwidth]{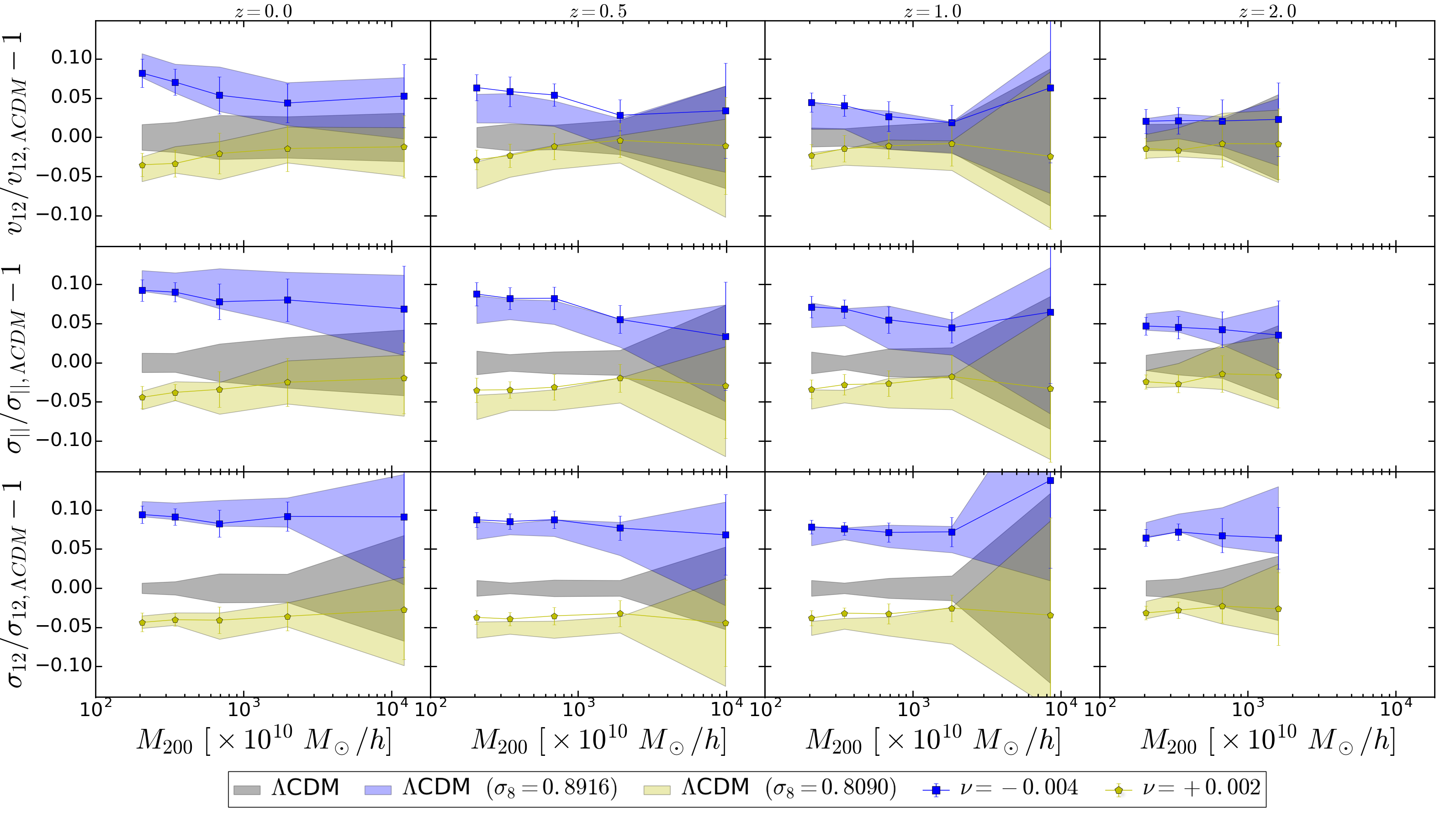}

\caption{ In the top row: the ratio of the pairwise velocity, $v_{12}$, between the R-FLRW simulations and the main $\Lambda$CDM run at fixed separation $r = 5$ Mpc/h for four different redshifts. In the second and third row: the same ratio calculated for $\sigma_{||}$ and $\sigma_{12}$ respectively. The error bars show the 1-$\sigma$ uncertainty calculated using a bootstrap technique. The same ratios are also calculated for two additional $\Lambda$CDM runs with the same $\sigma_8$ value at redshift $z=0$ of each R-FLRW simulation, represented as colored bands enclosing the 1-$\sigma$ region around each measurement.}
\label{fig:s8compar}
\end{figure*}

We focus on the pairwise velocity $v_{12}$ and its dispersions, $\sigma_{||}$ and $\sigma_{12}$, since these showed the largest deviations from $\Lambda$CDM in the previous section.
In the top row of Figure \ref{fig:s8compar} we plot the ratio of $v_{12}$ from the R-FLRW and the additional $\Lambda$CDM scenarios to the original $\Lambda$CDM simulation for four different redshifts. The next two rows similarly show the ratio of $\sigma_{||}$, and $\sigma_{12}$. We only consider one measurement at a separation of 5 Mpc/h for halos in five different mass bins, since this is the separation that provides the most robust determination while still having interesting deviations. Also, as we discussed before, the ratios show a negligible dependence on the separation.
Furthermore, to quantify the discriminatory power of these measurements, we calculate the intrinsic error of our method by bootstrapping our catalog to obtain 10 subsamples. The resulting variance is reported in Figure \ref{fig:s8compar} as 1-$\sigma$ error bars for the R-FLRW scenarios, and as 1-$\sigma$ error bands for the additional simulations. 

Overall, the additional $\Lambda$CDM simulations show trends which are very similar to the corresponding R-FLRW simulation, with the ratios being almost constant with mass, and roughly the same when compared with the main $\Lambda$CDM run from the original simulation suite. The first column in Figure \ref{fig:s8compar} presents the ratios at redshift $z=0$, where we can see a high degree of degeneracy between the R-FLRW scenarios and the corresponding $\Lambda$CDM simulations.
In \cite{Bibiano} we showed how this degeneracy could be broken by comparing the history of the various quantities under investigation, and thus in the remaining columns we show the history of each statistic at $z=0.5$, $z=1$ and $z=2$.
Here we find a shallow redshift dependence: at earlier redshifts the differences are slightly smaller, settling at around around 5\% and 2\% respectively for the $\nu < 0$  and $\nu > 0$ scenarios by redshift $z=2$.

The differences between the R-FLRW models and the main $\Lambda$CDM simulation are only significant (at least at the 3-$\sigma$ level) at low redshift. The differences between the R-FLRW models and the new $\Lambda$CDM simulations with the same $\sigma_8$ are not significant at any redshift.
This makes the degeneracy in the pairwise velocity statistics very difficult to disentangle from the determination of $\sigma_8$, even with a good knowledge of their redshift evolution.
This is because the pairwise velocity distribution is mainly affected by the new dynamics on linear and mildly non-linear scales, as we have shown in Figure \ref{fig:vonetwoparticles} where we investigated the the dark matter distribution. With our models it is only possible to break this degeneracy by probing fully non-linear scales and their redshift evolution, as we have shown in \cite{Bibiano}.

\section{Discussion}
\label{sec:discuss}
Thus far we have focused on the statistics of pairwise velocities to investigate how they are sensitive to the running of fundamental parameters introduced in the R-FLRW cosmology. This was motivated by the recent findings from \cite{hellwing} and \cite{ivarsen}, where  $\sigma_{||}$ and $\sigma_{12}$ were found to be particularly sensitive to modified gravity effects.
We developed our own high performance pairwise statistics code, and once run on our R-FLRW simulations found that such velocity  statistics are indeed modified by the new dynamics. In particular, the dispersions in the R-FLRW scenarios show excesses of up to 10\% when compared to \lcdm{}. 
Knowing that these excesses are usually degenerate with $\sigma_8$ we then ran additional simulations to quantify this. This revealed that the degree of degeneracy is significant. For example, the redshift $z=0$ pairwise velocity of halos in a R-FLRW scenario and a \lcdm{} universe with the same present-day $\sigma_8$ are indistinguishable to within the errors of our simulations. Unfortunately exploiting the redshift evolution of these observables does not help, as even at higher redshifts we fail to register any significant difference.
\begin{table}
\begin{center}
\begin{tabular}{lc}
\hhline{==}
Simulation & $\sigma_8(z=0)$ \\
\hline
$\Lambda$CDM &  0.809 \\
EXP001  & 0.825 \\
EXP002 & 0.875 \\
EXP003 & 0.967 \\
EXP008e3  & 0.895 \\
SUGRA003 & 0.806 \\

\hhline{==}
\end{tabular}
\caption{The different CoDECS simulations distinguished by the shape of the potential and the functional form and/or strength of the coupling between the dark energy field and dark matter. All the models have the same amplitude of matter perturbations at the redshift of the CMB the linear perturbations but evolve in a very different way, this results in very different values of $\sigma_8$ at redshift $z=0$, reported in the last column. }
\label{tab:codecs}

\end{center}
\end{table}

This $\sigma_8$ degeneracy was not studied previously in the literature when considering such observables. In the R-FLRW case it mostly results from the nature of the modification introduced by the model, having a time dependance but no scale dependence. 

To understand if this is a result shared by other dark energy models also making time dependent modifications to the standard cosmology we extend our analysis pipeline to examine the ``Coupled dark energy Cosmological Simulations'' (CoDECS) suite by \cite{baldiCodecs}. 
In such coupled dark energy models the dark energy component of the Universe is treated as a scalar field which evolves due to a potential, but also interacts with the dark matter fluid. This interaction results in a fifth force, felt only by dark matter particles, which influences the formation of structures both in the linear and non-linear regime.
For the CoDECS suite the models chosen are distinguished by the choice of potential for the dark energy scalar field, by the functional forms of the coupling, and within the same model by the strength of the coupling.
The names of the various runs, along with the value of $\sigma_8$ at redshift $z=0$, are reported in Table \ref{tab:codecs}: the EXP001, EXP002 and EXP003 runs simulate an exponential potential \citep{Wetterich88} and a constant coupling, with respectively increasing strength; the EXP008e3 run simulates an exponential potential with a variable coupling; and the SUGRA003 run simulates a supergravity potential \citep{Brax_1999} with a constant coupling. The details of the new dynamics and their background evolution is outside the scope of this work but is extensively reviewed in the introductory paper of the simulations suite, \cite{baldiCodecs}, to which we refer the interested reader.

While the additional parameters introduced by the different CoDECS models vary between the simulations, the base cosmological parameters are the same across all the runs and agree with the ``WMAP7 only Maximum Likelihood'' results by \cite{WMAP7}. The CoDECS suite was also the first to adopt a clear convention about the normalization of the initial conditions: all the models simulated agree on the measure of $\sigma_8$ at the redshift of recombination, $z \sim 1100$. This is the same convention we adopt for our simulations, and for all the models in this suite, except the SUGRA scenario, leads to a very different $\sigma_8$ at redshift $z=0$. For this reason we will focus our investigation of the pairwise velocity statistics on the degeneracies with the value of $\sigma_8$, as before. 

All the five simulations in CoDECS follow the evolution of $1024^3$ dark matter particles and $1024^3$ baryon particles in a periodic box 1 Gpc/h aside. The baryon particles are treated as collisionless and do not receive the usual hydrodynamic treatment. The necessity to include them as a separate kind of particle stems from the nature of the models, as the dark matter particles feel the additional fifth force --- which is not felt by the baryons --- due to the coupling being limited to the dark sector. Additionally, there is an energy-momentum transfer between the dark energy field and the dark matter fluid which results in an evolution of the dark matter particle mass with time. Aside from these additional details the main simulation algorithms and products are very similar to our simulations suite: the CoDECS N-body code is a modified version of the \textsc{gadget} code, and in post processing the FOF and SUBFIND algorithms were used to identify halos and sub-halos.

\begin{figure*}
\includegraphics[width=\textwidth]{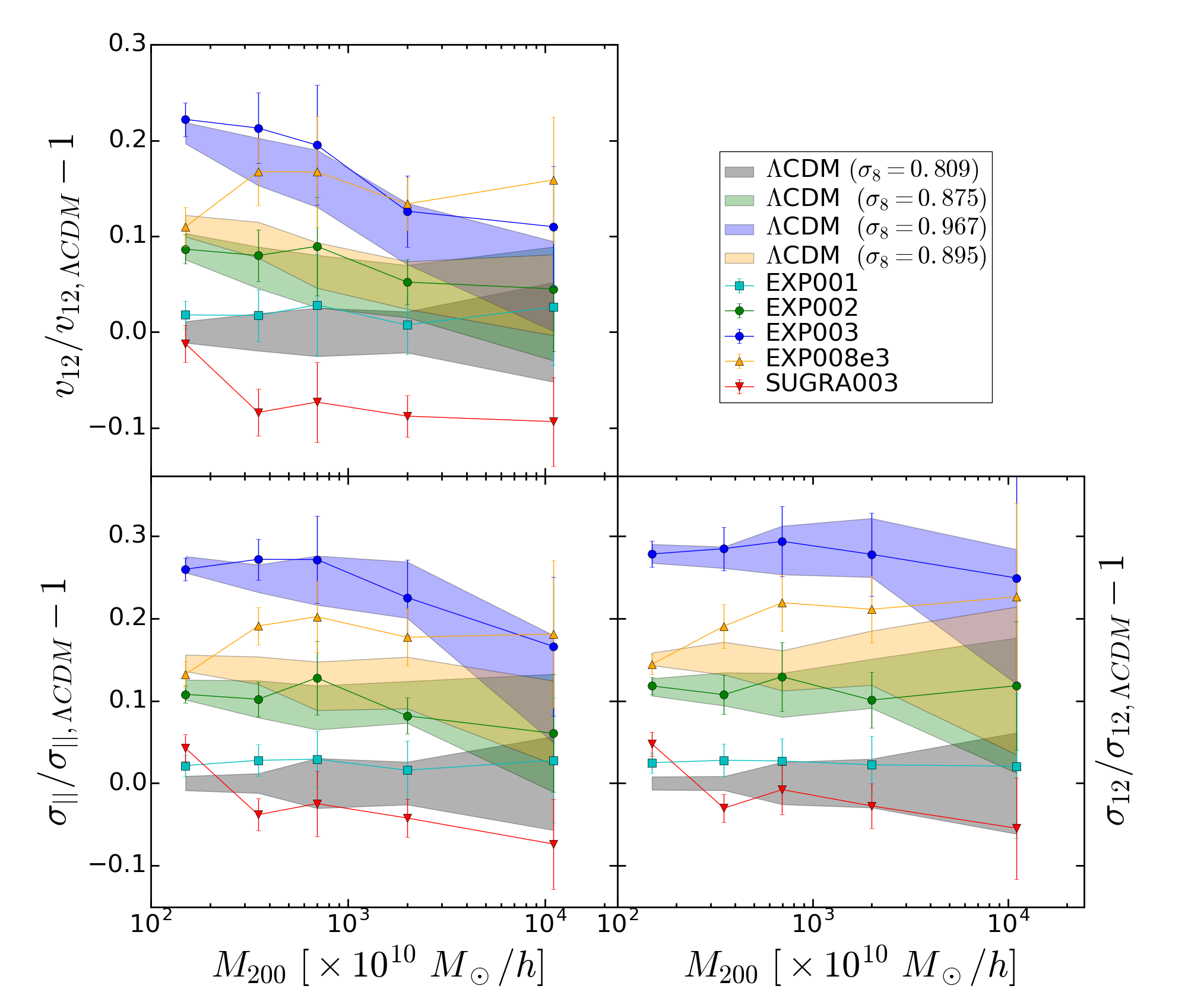}

\caption{ The ratio of the pairwise velocity, and the dispersions, $\sigma_{||}$ and $\sigma_{12}$ , between the CoDECS simulations and the main $\Lambda$CDM run at fixed separation $r = 5$ Mpc/h at redshift $z=0$. The error bars show the 1-$\sigma$ uncertainty calculated using a bootstrap technique. Each statistic is also calculated for three additional $\Lambda$CDM runs with the same $\sigma_8$ value at redshift $z=0$ as the CoDECS scenarios EXP002, EXP003 and EXP008e3, represented as colored bands enclosing the 1-$\sigma$ region around each measurement.}
\label{fig:codecs}
\end{figure*}

To begin, we run a partner set of \lcdm{} simulations having the same final $\sigma_8$ of each CoDECS simulation, as before.
We then adapt our analysis pipeline for the CoDECS simulation suite, splitting the full catalog into different mass bins and measuring the previous velocity statistics on all simulations:
$v_{12}$ and the dispersions $\sigma_{||}$ and $\sigma_{12}$, as well as performing the same error analysis described in Section \ref{sec:degeneracies}. 

In Figure \ref{fig:codecs} we present the results of this, where we find the same trends seen in the R-FLRW simulations, with both a mass and separation dependence. Here we focus on the magnitude of the dispersions which reveal some interesting deviations for two CoDECS simulations in particular, EXP003 and EXP008e3. These exhibit an excesses of between 15\% and 20\% in the pairwise velocity and between 20\% and 30\% in both dispersion measurements. The other simulations show smaller deviations from \lcdm{}, with the SUGRA model being the only one to exhibit smaller velocities and dispersion than the reference \lcdm{} simulation. Also, the SUGRA model presents a marked departure in its velocity measurement across all masses, although the dispersion values are difficult to distinguish from \lcdm{} within the errors of our analysis. The EXP001 model is completely degenerate with the reference \lcdm{} simulation across the measurements. 

To investigate the degeneracies found previously we compare our partner \lcdm{} simulations having the same value of $\sigma_8$ at redshift $z=0$ with each CoDECS simulation, shown by the colored bands, as labeled.
Here the $z=0$ measurements from the EXP003 and EXP002 simulations are now completely degenerate with the corresponding \lcdm{} simulation. We find the same result at higher redshift, not plotted for clarity, meaning we cannot use the redshift evolution to remove this degeneracy. The only scenario where the degeneracy is clearly broken is the EXP008e3 simulation with variable coupling, where we find around 5\% higher velocities and dispersions at redshift $z=0$ when compared with a \lcdm{} cosmology with the same $\sigma_8$. We also note another interesting feature in the $v_{12}$ measurement for the SUGRA scenario: in this model $\sigma_8$ at redshift $z=0$ is the closest to the reference \lcdm{} simulation out of all the models considered but produces larger deviations and a lower degree of degeneracy than the EXP001 scenario, even though the latter has a higher $\sigma_8$ at redshift $z=0$.

\refresponsestart{}Overall the $\sigma_8$ degeneracy substantially diminishes the discriminating power of such velocity observables when it comes to their application to most of the dark energy models presented here. Nevertheless, some scenarios, like the EXP008e3 and SUGRA simulations of the CoDECS suite, may still provide interesting ``smoking guns'', free of this degeneracy. A more detailed discussion of the differences between the CoDECS models can be found in \cite{Baldi:2010vv} and \cite{Baldi:2010vv2}, to which we refer the interested reader.\refresponseend{}
Such scenarios could reveal detectable signatures of velocity deviations from \lcdm{} if complemented with a very robust determination of $\sigma_8$ at redshift $z=0$.

It is the aim of both current and future peculiar velocity surveys to obtain such measurements. Pairwise velocity statistics from survey data have been attempted, e.g. \cite{Hawkins21112003}, and while the typical accuracy is usually in the tens of percent \citep{tullyvel}, future surveys aim to reduce this significantly by obtaining a larger number of objects covering a wider portion of the sky \citep{kodapec}. This will allow for accurate measurements of the growth rate and hence $\sigma_8$, and along with an improved precision in the velocities, result in the most accurate and discriminating pairwise velocity statistics to date.

The measurements we have presented here can be readily compared to observations provided that biases between the observed galaxy population and the simulated dark matter halo population \citep{Ferreirabias} are correctly accounted for, alongside a more complete treatment of baryonic effects. Such studies have been conducted in the past decade \citep{Kauffmann11021999}, and while simulations predict the presence of a clustering bias when comparing baryons and dark matter, they point to the absence of a velocity bias. This means that velocities measured from halos in cosmological simulations can be readily compared to galaxy observations without further adjustments, and so without the uncertainties that such adjustments always include.

\section{Summary}
\refresponsestart{}The aim of this work was to contribute to the current efforts being undertaken towards the exploitation of velocity statistics as probes of modified gravity and dark energy behaviours. In particular, we limited the scope of our study to pairwise velocities as these were recently shown to provide interesting signatures for modified gravity models. To this end we analyzed the velocities from a suite of N-body simulations in the R-FLRW cosmological model, extending the current literature and providing tests against recent cosmological observations and for future surveys. Our simulations use the latest determination of the cosmological parameters, and in their analysis we focused on the degeneracies that these observables present. We also extended our analysis to a second suite of simulations built on different dark energy models, increasing the number of scenarios with pairwise velocity predictions. By doing so we highlighted very important deviations which have the potential to be important ``smoking guns'' for some Coupled Dark Energy models. In fact, these modifications influence the velocity statistics in a way that is not degenerate with $\sigma_8$, meaning that the results presented here, in combination with future survey data, will help to further test and possibly falsify these scenarios. Our results warrant further theoretical study to better understand the mechanisms that lead to such differences in these models.\refresponseend{}

\section*{Acknowledgments}
We would like to thank Manodeep Sinha for making the Corrfunc code publicly available and for the helpful discussion in the development of our extension. We also thank Volker Springel for sharing the \textsc{gadget-3} code and Chris Blake for the discussion that spurred this study. We also thank the reviewer for their thorough review and highly appreciate the comments and suggestions, which significantly contributed to improving the quality of the paper.

\bibliography{bibliography/converted_to_latex.bib}

\end{document}